# Direct Energy-resolving CT Imaging via Energy-integrating CT images using a Unified Generative Adversarial Network

Lisha Yao[1,2], Sui Li[1,2], Manman Zhu[1,2], Dong Zeng[3,1], Zhaoying Bian[1,2], Jianhua Ma[1,2]

[1]Department of Biomedical Engineering, Southern Medical University, Guangzhou, Guangdong 510515, China

[2]Guangzhou Key Laboratory of Medical Radiation Imaging and Detection Technology, Guangzhou 510515, China

[3]College of Automation Science and Engineering, South China University of Technology, Guangzhou 510641, China

## ABSTRACT

Energy-resolving computed tomography (ErCT) has the ability to acquire energy-dependent measurements simultaneously and quantitative material information with improved contrast-to-noise ratio. Meanwhile, ErCT imaging system is usually equipped with an advanced photon counting detector, which is expensive and technically complex. Therefore, clinical ErCT scanners are not yet commercially available, and they are in various stage of completion. This makes the researchers less accessible to the ErCT images. In this work, we investigate to produce ErCT images directly from existing energy-integrating CT (EiCT) images via deep neural network. Specifically, different from other networks that produce ErCT images at one specific energy, this model employs a unified generative adversarial network (uGAN) to concurrently train EiCT datasets and ErCT datasets with different energies and then performs image-to-image translation from existing EiCT images to multiple ErCT image outputs at various energy bins. In this study, the present uGAN generates ErCT images at 70keV, 90keV, 110keV, and 130keV simultaneously from EiCT images at140kVp. We evaluate the present uGAN model on a set of over 1380 CT image slices and show that the present uGAN model can produce promising ErCT estimation results compared with the ground truth qualitatively and quantitatively.

## I. INTRODUCTION

Energy-resolving computed tomography (ErCT) can provide sufficient spectral information by simultaneously acquiring energy-dependent CT images. Different from traditional energy-integrating CT (EiCT), ErCT is equipped with an advanced photon counting detector. With this advanced technology, ErCT can improve contrast-to-noise ratio, decrease electronic noise, and improve dose efficiency, compared with EiCT [1]. Due to technical limitation, clinical ErCT scanners are not yet commercially available, and they are in various stage of completion [2]. Therefore, this makes researchers less accessible to the ErCT images compared to the EiCT images available.

Recently, deep learning has great potential in ErCT image generation or estimation due to its great computational power and fitting ability. For example, Li et al. designed a cascade deep convolutional network to estimate high-energy ErCT images from low-energy ErCT images [3]. Cong et al. proposed a deep-learning-based method to estimate ErCT images at specific energy from EiCT measurements to suppress beam-hardening artifacts [4]. Although these methods obtained acceptable ErCT images, it is limited that these models can only acquire ErCT images at one specific energy, which is far from the ErCT imaging requirement, i.e., multiple ErCT images at various energy bins.

Inspired by the work in [5], we investigate to produce ErCT images directly from existing EiCT images using a unified generative adversarial network (uGAN). This model is able to concurrently train EiCT datasets and ErCT datasets with different

Corresponding Author: Jianhua Ma (e-mail: jhma@smu.edu.cn).

energies and then performs image-to-image translation from existing EiCT images to multiple ErCT image outputs at various energy bins.

## II. METHODS

The goal of the proposed method is to simultaneously generate ErCT images in multiple energy bins from EiCT images. Fig. 1 demonstrates the framework of the proposed uGAN method, which consists of two modules, a generator $G$ and a discriminator $D$. $G$ is designed to translate an ErCT image into a fake ErCT (fErCT) image conditioned on the target label $c$ to fool D. Specifically, $c$ is generated randomly and $G$ can translate an input image into any target images. $D$ tries to distinguish between the fake image fErCT and the real ErCT (rErCT) image and classify the selective real images to its corresponding domain $c$. Additionally, an auxiliary classifier was induced to control multiple domains using a single discriminator, i.e., $D: x \rightarrow \{D_{src}(x), D_{cls}(x)\}$.

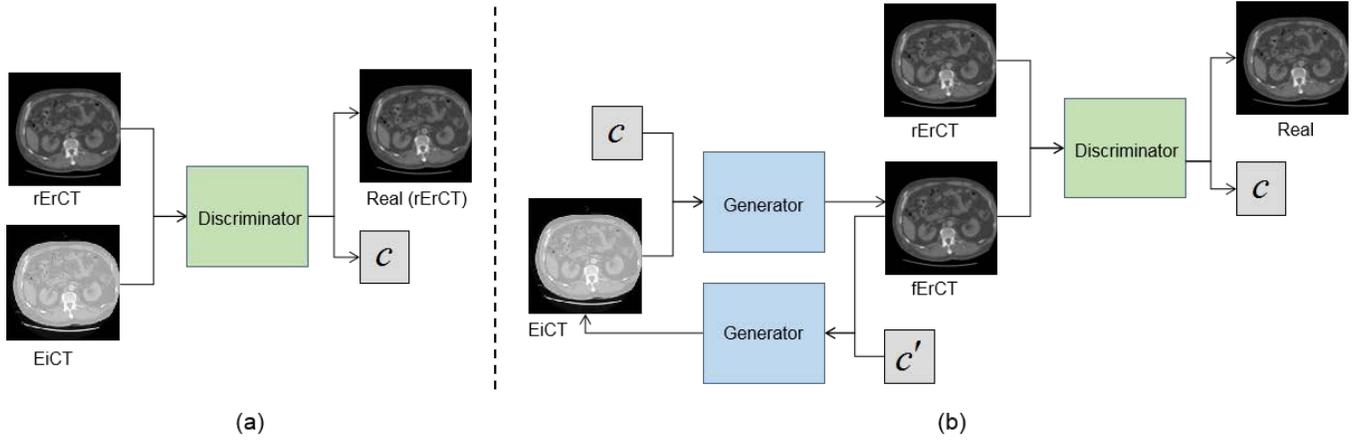

Fig. 1. Framework of the proposed method. (a) $D$ is trained to distinguish between the real ErCT image 'rErCT' and the fake image 'EiCT' and to classify the real image 'ErCT' to its corresponding target label. (b) $G$ tries to translate the EiCT image into a fake image 'fErCT' conditioned on the target label $c$. Similarly, $G$ tries to translate the fErCT image into the original EiCT image conditioned on the original label $c'$. Then, $D$ tries to distinguish between the real image 'rErCT' and the fake image 'fErCT' and to classify the selected real image to its corresponding target label $c$.

*A. Adversarial loss*

To generate ErCT images indistinguishable from real images, the adversarial loss can be expressed as:

$$\mathcal{L}_{adv} = \mathrm{E}_x[\log D_{src}(x)] + \mathrm{E}_{x,c}[\log(1-D_{src}(G(x,c)))], \qquad (1)$$

where $G$ aims to translate an EiCT image $x$ into an ErCT image conditioned on the target label domain $c$, i.e. $G(x,c)$, which tries to minimize the adversarial loss. $D_{src}$ aims to distinguish between the real and fake images, which tries to maximize the adversarial loss. $D$ and $G$ interact on each other until reaching the equilibrium.

*B. Domain classification loss*

To ensure that $G$ can translate an input image into a output image conditioned on the specific label $c$, we employ an auxiliary classifier $D_{cls}$ and decompose the domain classification loss into two terms:

$$\begin{aligned}\mathcal{L}_{cls}^{r} &= \mathrm{E}_{x,c'}[-\log D_{cls}(c'|x)], \\ \mathcal{L}_{cls}^{f} &= \mathrm{E}_{x,c}[-\log D_{cls}(c|G(x,c))],\end{aligned} \quad (2)$$

where $D_{cls}$ minimizes the domain classification loss of real images $\mathcal{L}_{cls}^{r}$ to guarantee that the input image $x$ can be classified to its original domain $c'$. And $G$ tries to minimize the domain classification loss of fake images $\mathcal{L}_{cls}^{f}$ that the output image $G(x,c)$ can be classified to its target domain $c$.

*C. Reconstruction loss*

The above two types of loss only guarantee that $G$ translates an input image into the target image with a specific label. However, the information of input images is inevitably lost during the translation process. To preserve the fidelity of input images, we induce a cycle consistency loss, which can be expressed as:

$$\mathcal{L}_{rec} = \mathrm{E}_{x,c,c'}[\|x - G(G(x,c),c')\|_{1}], \quad (3)$$

Specifically, we utilize L1 norm to regularize the loss of $x$ and $G(G(x,c),c')$.

Finally, the objective of the proposed method can be written as:

$$\begin{aligned}\mathcal{L}_{D} &= -\mathcal{L}_{adv} + \lambda_{cls}\mathcal{L}_{cls}^{r}, \\ \mathcal{L}_{G} &= \mathcal{L}_{adv} + \lambda_{cls}\mathcal{L}_{cls}^{f} + \lambda_{rec}\mathcal{L}_{rec},\end{aligned} \quad (4)$$

where $\lambda_{cls}$ and $\lambda_{rec}$ represent the hyper-parameters to control the weights of domain classification and reconstruction loss.

*D. Implementation*

We trained the network using the clinical patient images at 70keV, 90keV, 110keV, 130keV and 140kVp. Each energy bin contains 1384 slices collected from both chest and abdomen. Moreover, training samples at each energy bin are increased to 8304 slices by rotation and flip. Additional 130 slices at 140kVp were used for testing samples.

## III. RESULTS

Fig. 2 shows the ground-truth and the fErCT images obtained by the uGAN model at 70, 90, 110 and 130 keV, respectively. It can be observed that the present uGAN model yields the similar ErCT images with the ground truth in visual inspection. Moreover, the third row shows the difference images between the fErCT images and the ground truth. As shown, the difference is small except for high-intensity areas wherein the main reason is that the attenuation coefficient difference between the EiCT images and ErCT images is significant, and the network fails to fully learn the latent representation. Moreover, Fig. 3 shows the zoomed ROI images indicated by the red rectangle in Fig. 2. We can see that the present model can recover accurate structure details in low-intensity areas and miss some structure information in high-intensity areas. Tab. 1 lists the feature similarity (FSIM) values of the estimated fErCT images, and all of them are above 0.99, indicating that the present uGAN model can estimate accurate ErCT images from the EiCT images.

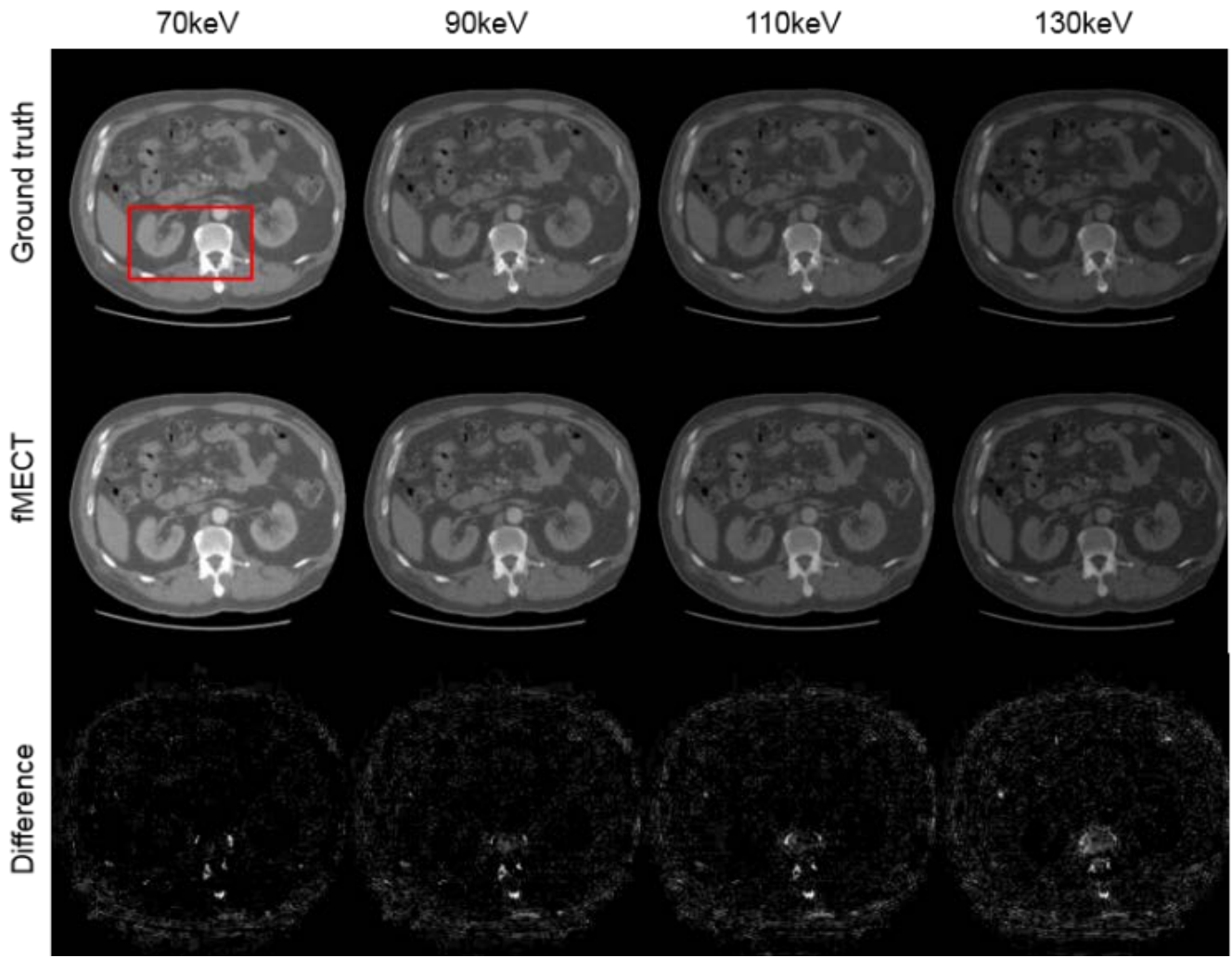

Fig. 2. The first row shows the ground truth, the second row shows the fErCT images, and the third row shows the difference images between the ground truth and the fErCT images.

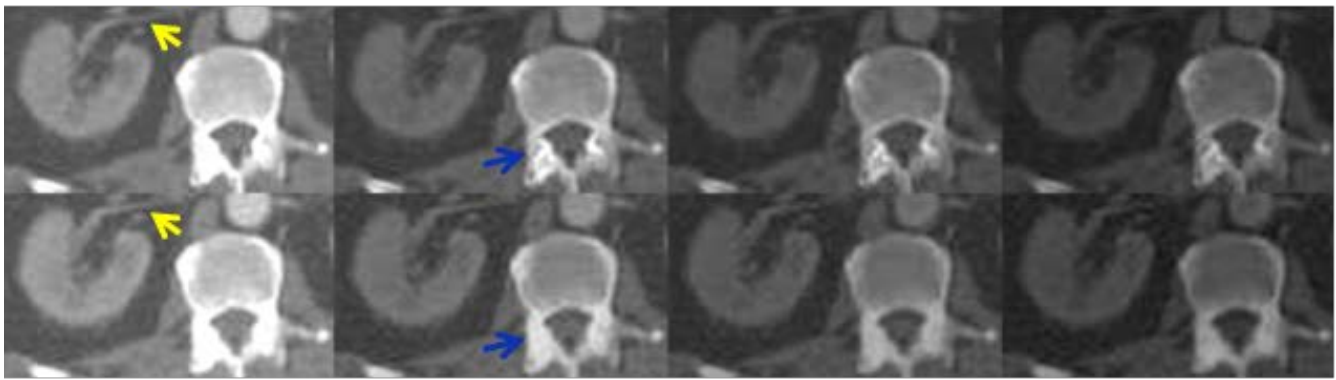

Fig. 3. Zoomed-in ROI images indicated by the red rectangle in Fig. 2.

Tab. 1 FSIM values of the pMECT images obtained by the proposed uGAN method.

|  | 70keV | 90keV | 110keV | 130keV |
|---|---|---|---|---|
| FSIM | 0.9979 | 0.9972 | 0.9961 | 0.9942 |

## IV. CONCLUSION

In this work, we present an uGAN model to produce ErCT images directly from existing EiCT images. Experimental results on clinical patient data demonstrate that the present uGAN model can yield high-quality and high-accuracy ErCT images. This provides a new way for ErCT image generation for research. However, it should be noted that the present uGAN model fails to reconstruct structure details in the high-intensity areas. In future study, we will try to add some other techniques to improve ErCT estimation performance, such as a prior-texture induced network [6].

## REFERENCES

[1] A. N. Primak, Giraldo, C. R. Juan, et al., "Dual-source dual-energy CT with additional tin filtration: dose and image quality evaluation in phantoms and in vivo Am J Roentgenol", 195(5): pp.1164-1174, 2010.

[2] M. J. Willemink, M. Persson, A. Pourmorteza, et al., "Photon-counting CT: technical principles and clinical prospects", Radiology, vol. 289, pp. 172656-172855, 2018.

[3] W. Cong, G. Wang, et al., "Monochromatic CT image reconstruction from current integrating raw data via deep learning", arXiv:1710.03784v2, 2017.

[4] S. Li, Y. Wang, Y. Liao, et al., "Pseudo dual energy CT imaging using deep learning based framework: initial study", arXiv:1711.07118v1, 2017.

[5] Y. Choi, M. Choi, M. Kim, et al., "StarGAN: unified generative adversarial networks for multi-domain image-to-image translation", arXiv:1711.09020v3, 2018.

[6] Z. Zhang, Z. Wang, Z. Lin, et al., "Image super-resolution by neural texture transfer", arXiv:1804.03360, 2019.